# Prototype of a front-end readout ASIC designed for the Water Cherenkov Detector Array in LHAASO

**Lei Zhao,**[a,b] **Weihao Wu,**[a,b] **Jianfeng Liu,**[a,b] **Yu Liang,**[a,b] **Jiajun Qin,**[a,b] **Li Yu,**[a,b] **Shubin Liu,**[a,b] **Qi An,**[a,b,*]

[a] *State Key Laboratory of Particle Detection and Electronics, University of Science and Technology of China, Hefei, 230026, China*

[b] *Department of Modern Physics, University of Science and Technology of China, Hefei, 230026, China*

*E-mail:* `anqi@ustc.edu.cn`

ABSTRACT: The Large High Altitude Air Shower Observatory is in the R&D phase, in which the Water Cherenkov Detector Array is an important part. The signals of Photo-Multiplier Tubes would vary from single photo electron to 4000 photo electrons, and both high precision charge and time measurement is required. To simplify the signal processing chain, the charge-to-time conversion method is employed. A prototype of the front-end readout ASIC is designed and fabricated in Chartered 0.35 μm CMOS technology, which integrates time disctrimination and converts the input charge information to pulse widths. With Time-to-Digital Converters, both time and charge can be digitized at the same time. We have conducted initial tests on this chip, and the results indicate that a time resolution better than 0.5 ns is achieved over the full dynamic range (1~ 4000 photo electrons, corresponding to 0.75 ~ 3000 pC with the threshold of 0.188 pC); the charge resolution is better than 1% with large input amplitudes (500 ~ 4000 photo electrons), and remains better than 15% with a 1 photo electron input amplitude, which is beyond the application requirement.

---

[*] Corresponding author.

# Contents

## 1. List of the abbreviations and acronyms

| | |
|---|---|
| Large High Altitude Air Shower Observatory: | LHAASO |
| Water Cherenkov Detector Array: | WCDA |
| Photo-Multiplier Tube: | PMT |
| Photo Electron: | P.E. |
| Charge-to-Time Converter: | QTC |
| Time-to-Digital Converter: | TDC |
| Field Programmable Gate Array: | FPGA |
| Process, Voltage & Temperature: | PVT |
| Monte Carlo: | MC |

## 2. Introduction

The project of Large High Altitude Air Shower Observatory (LHAASO) is in the R&D phase, which is proposed for a very high energy gamma ray source survey [1]. It is a complex consisting of several detector systems including a KM2A (1 km2 complex array), a WFCTA (Wide FOV Cherenkov Telescope Array), a SCDA (high threshold core-detector array) and a Water Cherenkov Detector Array (WCDA) [2]. As one of the kernel parts in LHAASO, the WCDA consists of four water ponds of 150 m×150 m, each with 900 Photo-Multiplier Tubes (PMTs) under water, as shown in figure 1.

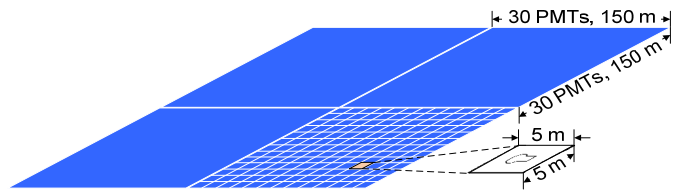

**Figure 1**. Structure of the WCDA in LHAASO.

The output signals of the PMTs are narrow pulses with a leading edge (10% to 90%) of 4 ns and a trailing edge of 16 ns, and the signal would vary from 1 photo electron (P.E.) to 4000 P.E. In this large dynamic range, both precise time and charge measurement is required [3], as listed in Table I.

| Parameter | Requirement |
|---|---|
| Channel number | 3600 |
| Signal dynamic range | 1 P.E. ~ 4000 P.E. |
| Resolution of charge measurement | 30% RMS@ 1 P.E.; 3% RMS@ 4000 P.E. |
| Bin size of time measurement | <1 ns |
| RMS of time measurement | <0.5 ns RMS |
| dynamic range of time measurement | 2 μs |

**Table I**. Design requirements of the WCDA readout electronics.

To simplify the system structure, a front-end readout ASIC is designed. Considering that both time and charge measurement is required, combination of Charge-to-Time Converters (QTCs) and Time-to-Digital Converters (TDCs) [4]–[10] is a preferred choice. Shown in figure 2 is the basic structure of front-end electronics for the LHAASO WCDA; the signal time is picked off through discriminators while the charge information is converted to a pulse width, and then the time and charge can be digitized simultaneously by a Field Programmable Gate Array (FPGA)-based TDC, which has become readily available [11]–[14] (in this system, an FPGA TDC with a bin size of around 0.3 ns based on multi-phase clock interpolation method [14] is adequate, the detailed information of which is not included in this paper). Therefore, the kernel parts are the QTC and discrimination circuits, which are integrated within one ASIC.

The main challenges in this front-end ASIC design include:

1) Large dynamic range. Input signal amplitude varies from 1 P.E. to 4000 P.E., and the discrimination threshold is required to be 1/4 P.E. The peaking current amplitude is 60 μA @ 1 P.E.; the corresponding input charge dynamic range is 0.75 ~ 3000 pC (with the threshold of 0.188 pC).

2) High measurement resolution. A charge resolution of 30% @ 1 P.E. and 3%@4000 P.E., as well as a time resolution better than 0.5 ns is required in the full dynamic range.

3) Good impedance matching. A 1~4000 dynamic range means that reflection of large signals would introduce interference on the measurement of small signals.

4) Low temperature drift. The front-end electronics are located in metal boxes above water of the WCDA. According to the thermal simulation results of the heat dissipation system, the extreme ambient temperature range of the ASIC is from 15 °C to 45 °C. Methods to reduce temperature drift are considered.

In this design, we implement two signal processing channels for one PMT with different gain factors to achieve a much larger dynamic range. Of course, with more channels, higher circuit complexity exists. Therefore, we should explore a structure to enhance the dynamic range and performance of each single channel, in order to minimize the channel number for each PMT. In this case, a current-mode processing structure [15], [16] is a favorable choice, considering its noise performance and simplicity.

For impedance matching, two basic ideas are considered: one is based on the equivalent impedance control of MOSFETs [17], and the other is termination using a 50 Ω resistor, which will be discussed in Section 3.

## 3. Circuits architecture

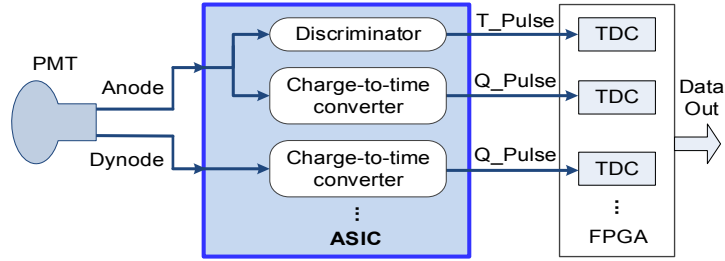

**Figure 2**. Structure of the PMT readout electronics in the WCDA of LHAASO.

As shown in figure 2, two signals are read out from each PMT: one is from the Anode, and the other is from a Dynode with a relatively lower gain. With this scheme, we can process the Anode signal to cover a signal range from 1 P.E. to 100 P.E., and use the Dynode signal to cover a range from 40 P.E. to 4000 P.E. for charge measurement. The Anode signal is also fed to discriminators for time measurement in the full scale range. The output pulse from the discriminator (marked as "T_Pulse") and the signal from the QTC circuit (marked as "Q_Pulse") are further digitized by the FPGA TDCs. To simplify the design and improve its performance, leading edge discrimination is employed and the time walk errors can be corrected using the charge measurement results. By integrating the front-end circuits within one ASIC, the system complexity can be greatly reduced.

### 3.1 Basic structure of the ASIC

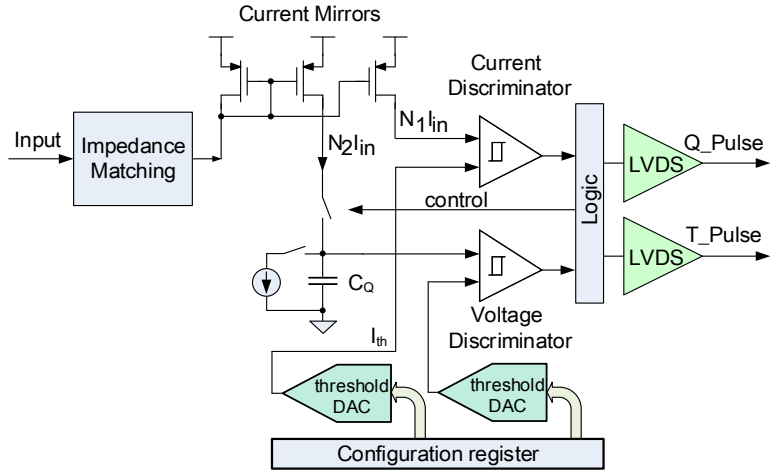

**Figure 3**. Block diagram of the ASIC.

Shown in figure 3 is the basic structure of the ASIC. As aforementioned, the large dynamic range requires good impedance matching at the input port. The input signal is afterwards split into two paths through a current mirror: one is imported to a current discriminator for time measurement, while the other is injected into a capacitor ($C_Q$ in figure 3) for charge measurement. The switches in figure 3 are controlled by a combinatory logic circuit to implement the charging and discharging process.

### 3.2 Input stage

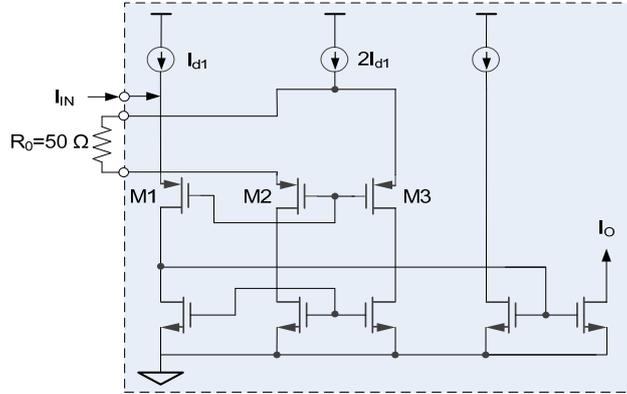

**Figure 4**. Impedance matching circuit with MOSFETs.

Considering that the signal from the PMT is in a current mode, and the QTC circuit is also based on current signal manipulation, the simplest method is to control the equivalent impedance of the MOSFET to achieve a 50 Ω input impedance. We designed impedance matching circuit inspired by the structure in NINO ASIC [18], as shown in figure 4. However, according to our simulation and test results (as shown in figure 5), the equivalent impedance of the circuit would vary with different amplitudes, and thus precise impedance matching cannot be achieved. It is probably due to the equivalent impedance variation with input amplitudes, which could also be susceptible to the PVT (Process, Voltage & Temperature) variation (such as film thickness, lateral dimensions, and doping concentrations [19] during the ASIC manufacturing process).

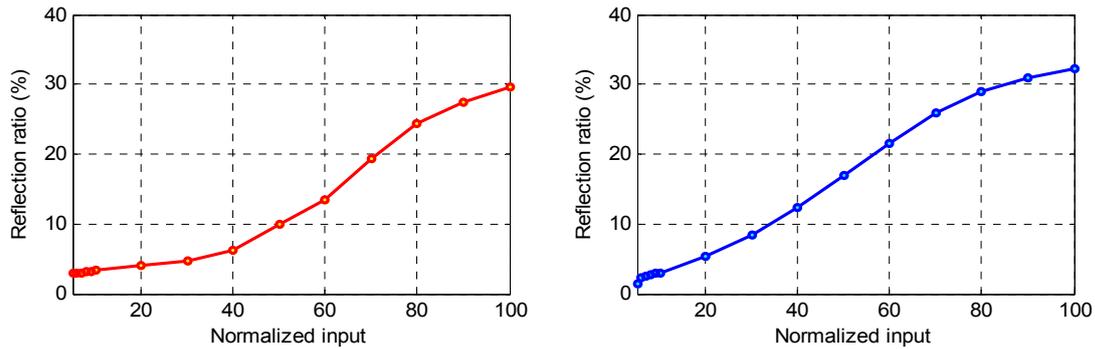

**Figure 5**. (L) Simulation results of reflection ratio for different input signal amplitudes. (R) Test results of reflection ratio.

Considering the limitation of the impedance matching circuit using MOSFETs, we implement the signal termination using a 50 Ω resistor, as shown in figure 6.

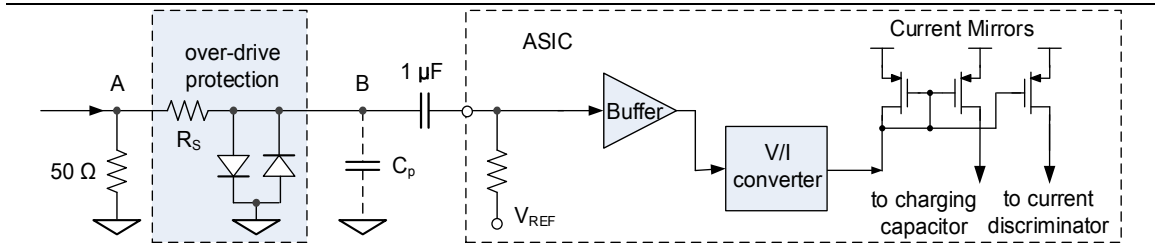

**Figure 6**. Impedance matching using a 50 Ω resistor.

The input signal is terminated with an off-chip 50 Ω resistor at the input of the ASIC. Since the input current is converted to a voltage signal simultaneously, a Voltage-to-Current (V/I) converter is designed to convert the signal back to a current, which is processed by the subsequent current-mode QTC circuit. With this scheme, no obvious signal reflection was observed in laboratory tests.

To protect the ASIC, over-drive protection circuit should be considered. In this design, we use diodes with a series resistor ($R_S$) to achieve this, as shown in figure 6.

### 3.3 V/I conversion circuit

The current-mode signal processing has the advantage of compact structure, less sensitivity to the fluctuation of supply voltage, higher pole frequency and so on [20]. Since the input current is converted to a voltage signal by the input 50 Ω resistor, V/I conversion [21]–[23] is implemented in order to perform the current-mode QTC processing. Shown in figure 7 is the V/I conversion circuit based on the FVF_RCM (flipped voltage follower regulated cascode mirror) structure, which features a big input swing and low quiescent current. With $R_1$=750 Ω and a voltage swing of $V_{IN}$ more than 1 V (the output current varies from 13 μA to 1.3 mA), an input range of more than 1 P.E. to 100 P.E. is covered. To reduce transconductance variation of the V/I conversion with different temperatures, $R_1$ in figure 7 is designed by combination of "npolyf_u" and "npoly_u_1k" in Chartered 0.35 μm CMOS technology, which have opposite temperature coefficients to achieve an approximately constant resistance of $R_1$. Simulation results indicate that the resistance of $R_1$ varies within ±0.8% over the temperature range from 15 °C to 45 °C.

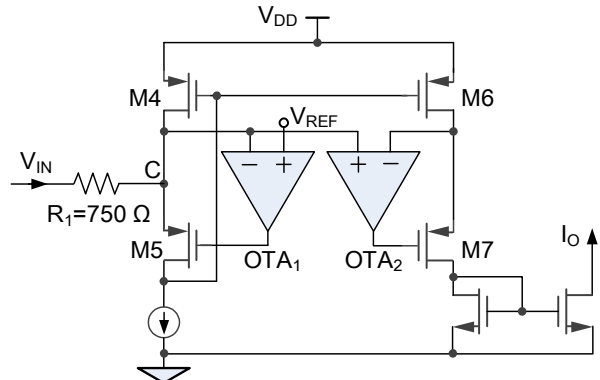

**Figure 7**. Voltage-to-Current conversion circuit.

## 3.4 QTC circuit design

The kernel part of the front-end electronics is to implement the QTC circuit. As shown in figure 8, the signal from the V/I converter is split into three paths. One is charged on a capacitor $C_Q$ (10 pF) for charge measurement, and the other two are fed to two current discriminators with a low (1/4 P.E.) and a higher threshold (3 P.E. in the test, user controlled), respectively. The reason for employing this two-threshold discrimination is to avoid deterioration of time measurement resolution which would be caused by noise or interference in the baseline of the large input signal.

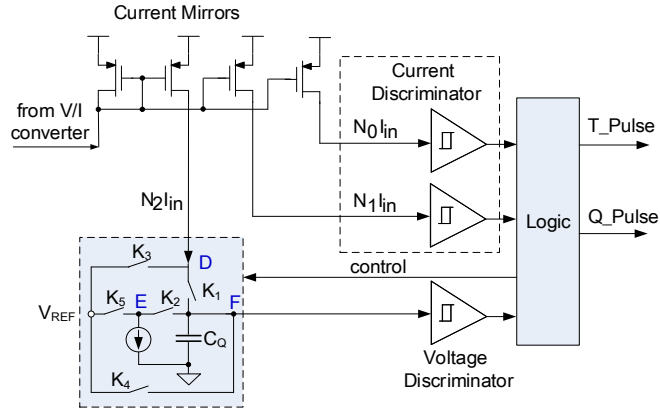

**Figure 8**. Block diagram of the QTC circuit.

One of the outputs from these two discriminators is selected as the final time result according to the charge measurement results. The output signal from the low-threshold current discriminator is also used to control the charging process for $C_Q$. When the charging process finishes, a current source of 30 μA is employed to discharge the capacitor. The voltage on $C_Q$ is compared with a certain threshold by a voltage discriminator, and thus the QTC process is accomplished. Besides the switches $K_1$ and $K_2$ used in the charging & discharging process, the other three switches are employed to reset the nodes D, E, and F to $V_{REF}$ (0.5 V) in a specified sequence, in order to make the circuit function well.

## 3.5 Time discrimination

The current discriminator [24]–[27] is one of the most important parts in this ASIC. A fast current discriminator structure [28] is employed in the time discrimination circuit, as shown in figure 9. By adjusting the sizes of the MOSFETs, a delay less than 1 ns is achieved with a step signal in the simulation. $C_A$ in figure 9 is used for AC coupling and thus filtering out the current baseline of the signal that could be susceptible to PVT errors. The $PM_0$ functions as a diode, preventing over saturated state of the circuit.

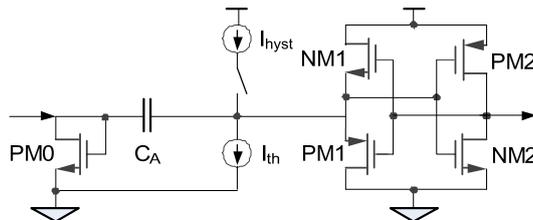

**Figure 9**. Time discrimination circuit.

Shown in figure 10 are the Monte Carlo (MC) simulations results of the baseline of the discriminator input current, which indicate that the baseline variation is greatly reduced with the AC coupling.

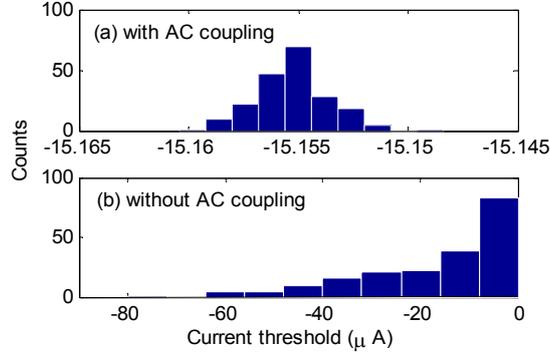

**Figure 10**. MC simulation results of the current baseline.

The output of this discriminator is distributed to two paths: one is driven to a Low-Voltage Differential Signaling (LVDS) [29] buffer as the output "T_Pulse" (in figure 8) for time measurement, and the other is used to control the charging process of the capacitor $C_Q$ in figure 3 and figure 8.

## 4. Simulation

We conducted a series of simulations to evaluate performance of the ASIC before actual fabrication. In the simulation, we used the waveform of the PMT (R5912 [30]) output signal as the signal source. To maximally approximate the real application situation, we added a 30-meter transmission line model and all the front-end coupling circuits in figure 6.

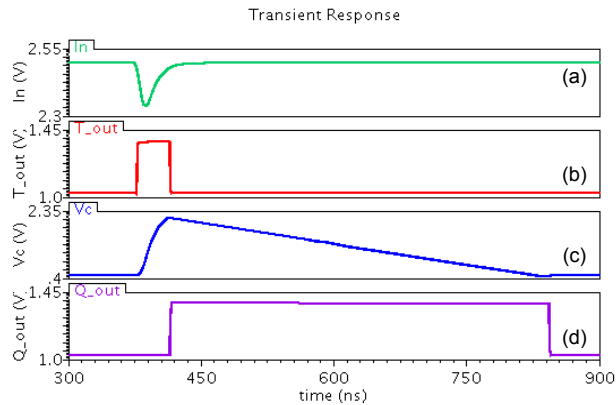

**Figure 11**. Waveform simulation results. The waveforms of (a) to (d) correspond to the input signal of the ASIC, the output of the current discriminator, the signal on the integration capacitor, and the QTC output signal (i.e. "Q_Pulse" in figure 8).

To confirm the functionality of this ASIC, we conducted transient simulations. Shown in figure 11 are the typical waveform simulation results of kernel nodes within the circuits. The charging and discharging processes can be clearly observed in the waveforms.

Then, we conducted a series of MC simulations to evaluate the circuit performance by sweeping the input signal amplitude. By obtaining the current noise and slope of the input signal of the current discriminator through noise and transient simulations, the output signal jitter (i.e. the time resolution) can be calculated. As shown in figure 12, with a 30-meter cable ahead of this ASIC, a time resolution better than 500 ps is achieved, and a time walk less than 20 ns can be guaranteed.

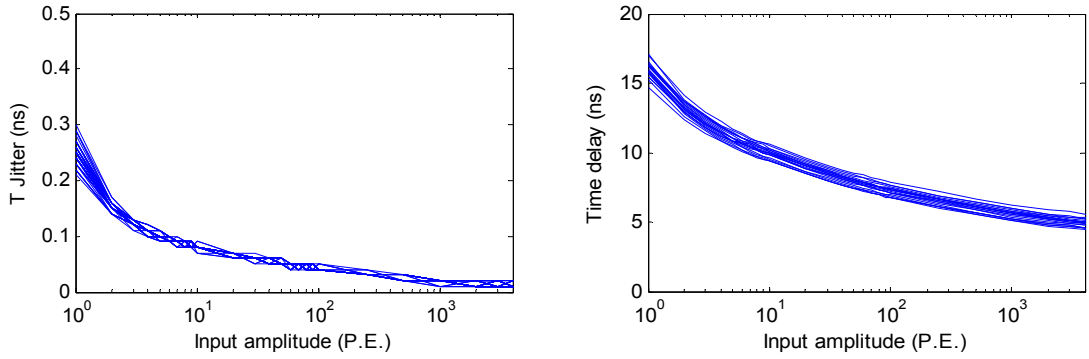

**Figure 12**. (L) Time resolution simulation results. (R) Simulation results of the time delay (referring to the output delay from the signal beginning) vs. input amplitude.

Shown in figure 13 (L) are the pulse width simulation results of the QTC output signal ('Q_Pulse' in figure 8) with different input signal amplitudes; a dead time less than 500 ns can be achieved for charge measurement.

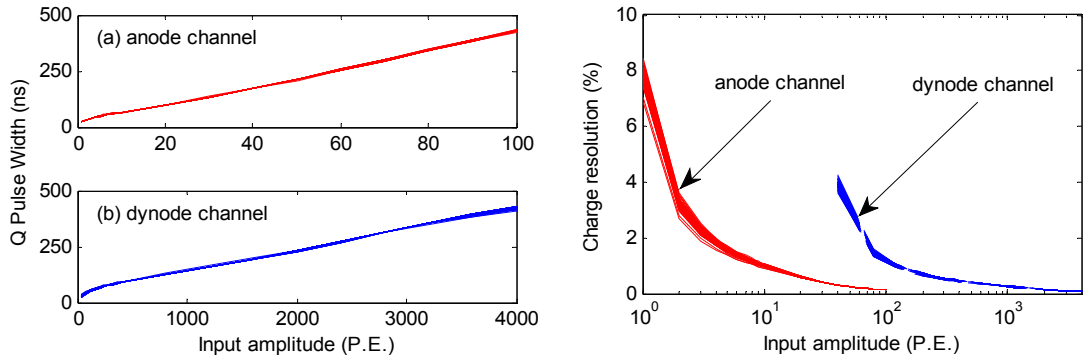

**Figure 13**. (L) QTC output signal widths with different input amplitudes. (R) Charge resolution simulation results.

By estimating through simulation the noise of different components and the slope of the signal on the integration capacitor $C_Q$ crossing the threshold of the voltage discriminator in figure 8, the RMS of the QTC output pulse width can be calculated. With the curve in figure 13 (L), the charge measurement resolution can be further estimated. The simulation results are shown in figure 13 (R), which indicate a charge resolution better than 10% can be achieved in the input amplitude range from 1 P.E. to 4000 P.E.

## 5. Circuits Performance

Shown in figure 14 is the layout of the prototype ASIC, which is designed in the Chartered 0.35 μm CMOS technology. Four channels (corresponding to two PMT readout channels) are integrated in a 3 mm × 3 mm block. After confirming the basic functionality and performance through this prototype, we will integrate more channels within one chip in the following version.

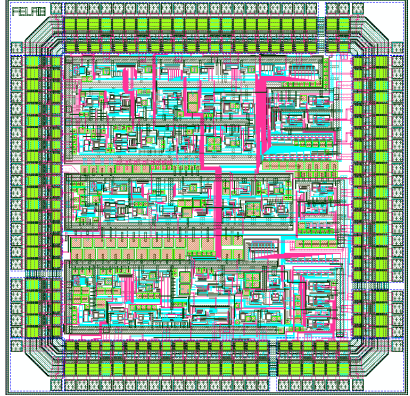

**Figure 14**. ASIC layout.

After ASIC fabrication and packaging, we conducted tests in the laboratory. As shown in figure 15, we use a signal source (Agilent 81160A) to generate the input signal according to the signal waveform of the PMT (R5912). A 30-meter cable is connected between the signal generator and the ASIC test board. We changed the input amplitudes with an attenuator (WAVETEK 5080.1) and conducted a series of tests to evaluate the performance of this prototype ASIC circuit.

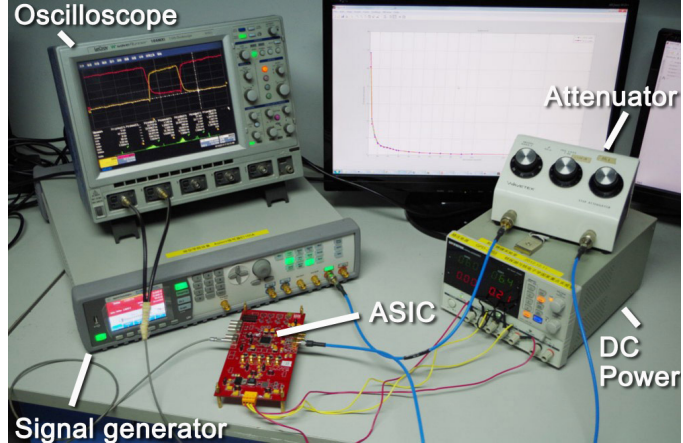

**Figure 15**. ASIC test system.

Shown in figure 16 are the waveforms of kernel nodes, which are driven out to special test pins by the buffers inside the ASIC. The waveforms in figure 16 concord well with those in figure 11, which indicates that this prototype ASIC functions well.

– 9 –

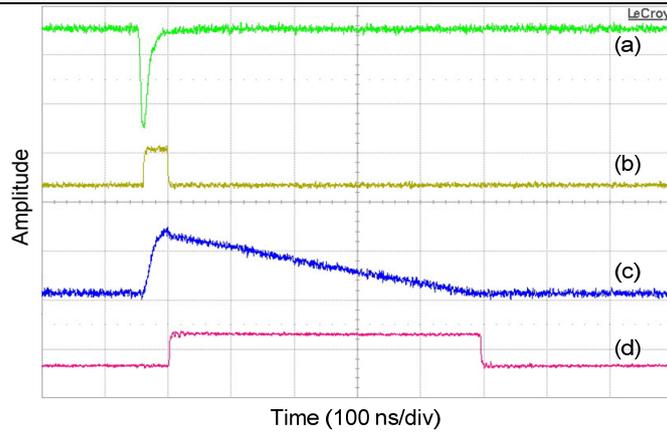

**Figure 16**. Waveform test results. The waveforms of (a) to (d) correspond to the input signal, the output of the current discriminator, the signal on the integration capacitor, and the QTC output signal (i.e. "Q_Pulse" in figure 8), and the amplitude scale is 100 mV/div, 500 mV/div, 1 V/div, and 500 mV/div, respectively.

We also conducted tests to evaluate the time and charge measurement performance of this ASIC, and a total of six pairs of anode and dynode channels were tested. The test of the time measurement performance is based on the "cable-delay test" method [6], [10], [31]. As shown in figure 17, the jitter of the 'T_Pulse' in figure 8 (i.e. the time resolution) is better than 400 ps, and a time walk less than 15 ns is observed.

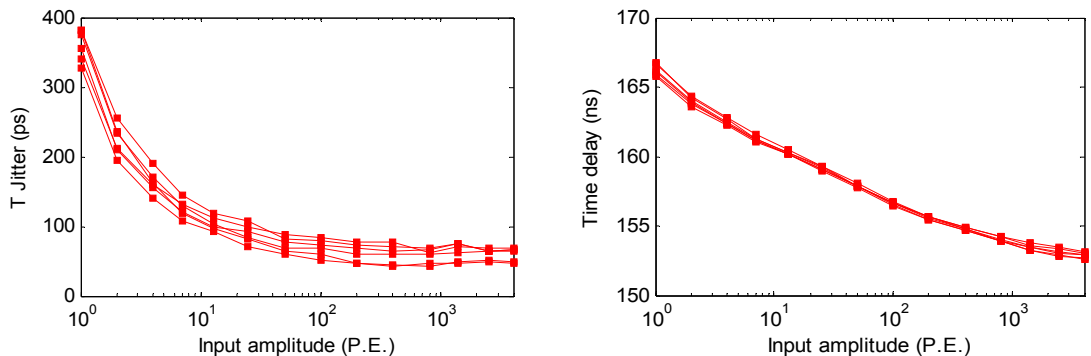

**Figure 17**. (L) Time resolution test results. (R) Test results of the time delay vs. input amplitude.

Shown in figure 18 (L) are the QTC output signal width test results with different input amplitudes, and the charge measurement resolution is shown in figure 18 (R), which indicates a total dynamic range of more than 4000 can be achieved, and the charge resolution is better than 15% at 1 P.E. and better than 1% with large input amplitudes (500 ~ 4000 P.E.).

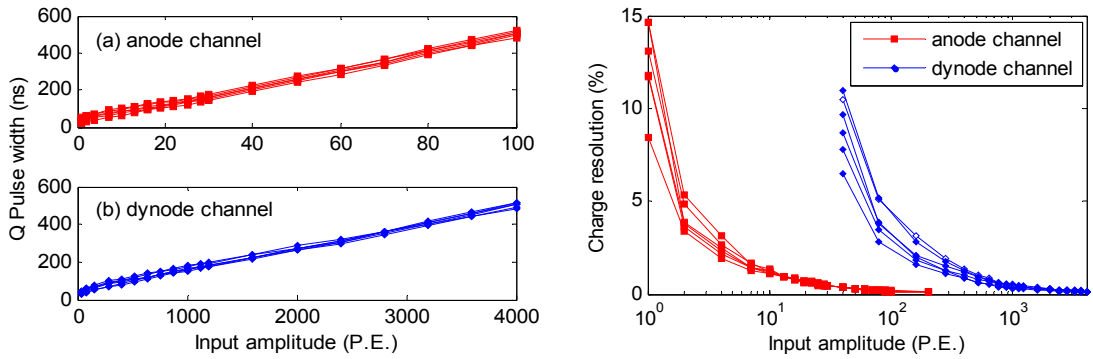

**Figure 18**. (L) Test results of the QTC output signal widths with different input amplitudes. (R) Charge resolution test results.

Besides, we have also conducted initial temperature drift tests over a dynamic range from 0 °C to 50 °C. The temperature drift of this ASIC mainly comes from variation of the resistor in the V/I conversion and gain of the current mirror with different temperatures. Considerations to reduce the temperature drift effect include low temperature coefficient resistor design (in figure 7), AC coupling before the current discriminator (in figure 9), as well as dummy circuits design of the V/I converter and the current mirror to further reduce the temperature drift. The test results indicate that the output width of "Q_Pulse" varies within ±5%, and the charge resolution stays better than 16% at 1 P.E., 1% with large input amplitudes, with time resolution better than 0.4 ns. And the temperature drift performance would be further enhanced with calibration and correction versus temperature in system level.

## 6. Conclusion

We present the design of a prototype front-end ASIC in the readout electronics of the WCDA in LHAASO. Based on the current-mode QTC processing, a good system simplicity can be attained. With multiple processing channels for one PMT with different gain factors, a large dynamic range is achieved. We also conducted initial tests on the prototype ASIC, and the results indicate that a time resolution better than 400 ps is achieved, and the charge resolution is better than 15% at 1 P.E. and remains better than 1% with large input amplitudes, beyond the requirement.

## Acknowledgments

The authors would like to thank Allan K. Lan of the MD Anderson Cancer Center in University of Texas for his help regarding this work over the years.